\documentclass[12pt, prd, showpacs,showkeys,superscriptaddress,nofootinbib]{revtex4}

\usepackage{amsmath,amssymb}

\usepackage{color}
\definecolor{DarkGreen}{rgb}{0,0.4,0}
\definecolor{DeepBlue}{rgb}{0,0,.4}
\usepackage[naturalnames=true,colorlinks=true,
citecolor=DarkGreen,urlcolor=blue,linkcolor=DeepBlue]{hyperref}

\begin{document}

\title{Propagation of electromagnetic waves\\
in Bose-Einstein condensate of atoms with dipole moments}
\author{Yu.M. Poluektov}
\email{yuripoluektov@kipt.kharkov.ua}
\affiliation{Kharkov Institute of Physics and Technology,\\
1 Akademicheskaya, Kharkov 61108, Ukraine}
\author{I. V. Tanatarov}
\affiliation{Kharkov Institute of Physics and Technology,\\
1 Akademicheskaya, Kharkov 61108, Ukraine}
\affiliation{Department of Physics and Technology, Kharkov V.N. Karazin National
University, 4 Svoboda Square, Kharkov 61022, Ukraine}

\begin{abstract}
We study the propagation of electromagnetic waves in the Bose-Einstein condensate of atoms with both intrinsic dipole moments and those induced by the electric field. The modified Gross--Pitaevskii equation is used, which takes into account relaxation and interaction with the electromagnetic field. Two cases are considered: 1) when the dispersion curves of the electromagnetic wave and of the condensate excitations do not intercross and 2) when the condensate excitations' spectrum has a gap and the two dispersion curves do intercross. In the second case the two branches hybridize. It is shown that propagation of sound waves can be accompanied by oscillation of the electromagnetic field. The impact is studied of the  dipole-dipole interaction on the character of electromagnetic and acoustic waves' propagation in the Bose-Einstein condensate.
\end{abstract}


\maketitle
\newpage \tableofcontents \newpage

\section{Introduction}
Recently, considerable attention is being paid to the behavior of superfluid Bose systems in electric and magnetic fields. The interest in these issues is stimulated by experimental studies of Bose-Einstein condensates (BEC) in the traps generated by either magnetic or laser fields \cite{Pit}. Studies of propagation of light in atomic gases \cite{Hau,Liu} imply the peculiar character of interaction of electromagnetic field with the many-body Bose systems composed of neutral atoms. The series of experimental studies \cite{Ryb04}-\cite{Ryb10} report unexpectedly high electric activity of superfluid helium exhibited under different conditions. In order to understand the effects observed in these and similar experiments one should investigate in detail the interaction of electromagnetic field with a many-particle system of Bose atoms in coherent state.

The interaction of electromagnetic field with a system of electric charges is realized through the multipole moments of the system. If the system is electrically neutral, the next most important characteristic that describes its interaction with the electric field is its dipole moment. There are arguments in favor of the conjecture that a helium atom, which in the free state has no intrinsic dipole moment, in liquid helium may spontaneously acquire the intrinsic dipole moment \cite{YM13}. Therefore it is important to carry out the detailed theoretical study of the properties of superfluid system of atoms that have an intrinsic dipole moment. This will allow one to compare the theoretical predictions with the phenomena observed in the experiments. Propagation of electromagnetic waves in the BEC taking into account the internal structure of the atoms in the ideal gas model was studied in \cite{Morice,Ruostekoski,SlSot} and the approach accounting for the structure of atoms in the framework of the modified Gross-Pitaevskii (GP) was proposed in \cite{YM11}.

In this paper we study the propagation of electromagnetic waves in an anisotropic superfluid system of atoms with intrinsic dipole moments using the modified GP equation, which takes into account BEC relaxation. Due to the assumed presence of intrinsic atomic dipole moments, the electromagnetic and sound waves in such a medium are coupled. In the second section we modify the nonstationary GP equation to incorporate relaxation processes in the condensate by introducing a phenomenological dissipative coefficient, which determines the third coefficient of viscosity and the time of homogeneous relaxation. Interaction of atoms in the condensate with the electric field in the dipole approximation is introduced in the third section, taking into account both the atom's intrinsic and induced dipole moments. In the fourth section we calculate the permittivity of atomic condensate, taking into account the short-range interparticle interaction forces. It is shown that the BEC of atoms with dipole moments is a medium with both temporal and spatial dispersion. The propagation of electromagnetic and acoustic waves in the condensate is studied in section five. The cases when the wave propagates along and perpendicular to the orientation of the dipole moments are considered separately, taking into account dissipation. We draw attention to the fact that the propagation of sound waves in the condensate is coupled to electric field oscillations. In section six we study propagation of electromagnetic waves in the case when the spectrum of the superfluid system's quasiparticles has an energy gap. In this case, the dispersion curves of such excitations and electromagnetic waves intercross and we observe the branches' hybridization. The final seventh section is devoted to investigation of the effects of long-range dipole-dipole interaction on the modified GP equation and on the properties of both acoustic and electromagnetic waves.

\section{Dissipation in the Gross-Pitaevskii approach}
The dynamic GP equation for the macroscopic condensate wave function $\psi =\psi (\mathbf{r},t)$, which interacts with external electromagnetic field $U(\mathbf{r},t)$ \cite{Pit} 
\begin{equation}
\label{EQn_1_}
i\hbar \dot{\psi }=-\frac{\hbar ^{2} }{2m} \Delta \psi +\left[U\left(\mathbf{r},t\right)-\mu \right]\psi +g\left|\psi \right|^{2} \psi , 
\end{equation}
can be derived in the Lagrangian formalism \cite{Goldstein,YM03} by choosing the Lagrange function in the form
\begin{equation}
\label{EQn_2_}
\Lambda =i\frac{\hbar }{2} \left(\psi ^{*} \dot{\psi }-\dot{\psi }^{*} \psi \right)
	-\frac{\hbar ^{2} }{2m} \left|\nabla \psi \right|^{2} 
		-\big(U\left(\mathbf{r},t\right)-\mu \big)\left|\psi \right|^{2} 
		-\frac{g}{2} \left|\psi \right|^{4} .
\end{equation}
Here $\mu $ is chemical potential, which can be expressed through the total number of particles through the relation
\begin{equation}
\label{EQn_3_}
	N=\int \left|\psi _{0} \right|^{2} d\mathbf{r} ,
\end{equation}
where $\psi_0$ is the equilibrium macroscopic wave function. The equation \eqref{EQn_1_} is time-reversible, i.e. invariant under transformation $t\mapsto -t,\quad \psi \mapsto \psi^*$, and describes the dynamics of the condensate neglecting any possible dissipative processes. It also implies conservation of the total number of condensate particles. It is obvious, that in non-stationary processes the condensate particles can pass to excited quasiparticle states and the number of particles is not conserved. If at the given time the system is in a non-equilibrium state, in which a portion of the particles is in the one-particle condensate and a portion forms the gas of quasiparticles, this state will relax to the equilibrium one and in time all particles will become part of the condensate. The attenuation of oscillations of the atomic condensate in magnetic traps was observed experimentally \cite{Mewes} and turned out to be small.

The dissipative processes in one-particle condensate can be taken into account phenomenologically in the frame of Lagrange formalism by introducing the dissipative function \cite{Goldstein}. In this case the Euler-Lagrange equations take the form
\begin{equation}
\frac{d}{d{\kern 1pt} t} \frac{\partial \Lambda }{\partial \dot{\psi }} -\frac{\partial \Lambda }{\partial \psi } +\nabla \frac{\partial \Lambda }{\partial \nabla \psi } 
	=-\frac{\partial {\rm D}}{\partial \dot{\psi }} , 
\label{EQn_4_}
\end{equation}
where ${\rm D}$ is the dissipative function, usually chosen to be quadratic in velocity \cite{LL}
\begin{equation}
{\rm D}=\hbar \gamma \dot{\psi }^{*} \dot{\psi },
 \label{EQn_5_}
\end{equation}
while $\gamma $ is the phenomenological dimensionless dissipative coefficient. Then, taking into account \eqref{EQn_2_}, one derives the equation of motion, which only differs from \eqref{EQn_1_} by the term with time derivative and real pre-factor $\sim \gamma$:
\begin{equation}
i\hbar \dot{\psi }=-\frac{\hbar ^{2} }{2m} \Delta \psi 
	+\big(U\left(\mathbf{r},t\right)-\mu \big)\psi 
	+g\left|\psi \right|^{2} \psi +\hbar \gamma \dot{\psi }.
 \label{EQn_6_}
 \end{equation}
Formally it can be obtained from  \eqref{EQn_1_} by substitution $i\mapsto i-\gamma $. 

Let us make a remark regarding the description of dissipative processes in this approach. 
The standard GP equation is used to describe the condensate at zero temperature. Dissipative processes are accompanied by the production of entropy and heat, and thus an increase in temperature. In the approach used here these temperature effects are not accounted for, and it is assumed that, as in the standard case, the system is at zero temperature. Practically, this means that the generated heat is removed from the system fast enough so that its temperature does not have time to change appreciably. Such an approximate allowance for dissipation is fully equivalent to the description of friction in mechanics, where the dissipative function describes the transition of mechanical energy into heat, but the concepts of temperature and entropy are not used.

It can be shown \cite{YM14}, that the dimensionless dissipative coefficient $\gamma$ is related to the coefficient of third viscosity $\zeta_3$ and the time of homogeneous relaxation of particle density in the condensate $\tau_0$ by the relations
\begin{equation}
\zeta _{3} =\frac{\hbar \gamma }{2m^{2} n_{0} } ,\quad \quad \quad \quad \tau _{0} =\frac{\hbar }{2g\gamma {\kern 1pt} n_{0} } .
\label{EQn_7_}
\end{equation}

Estimates show \cite{YM14}, that $\gamma \sim 10^{-6} \doteq 10^{-5} $. This coefficient is equal to the inverse quality factor of the system.

\section{Interaction of condensate with electromagnetic field}
The potential energy of a condensate atom's interaction with the electromagnetic field in the dipole approximation is
\begin{equation}
U\left(\mathbf{r},t\right)=-\mathbf{d}\cdot \mathbf{E}(\mathbf{r},t) , 
\label{EQn_8_}
\end{equation}
where $\mathbf{d}$ is the atom's dipole moment, and \[\mathbf{E}\left(\mathbf{r},t\right)=\mathbf{E}_{0} +\delta \mathbf{E}\left(\mathbf{r},t\right)\]
is the electric field acting on it, which consists of the constant and variable parts. The local electric field acting on an atom in dielectric is different from the external field \cite{Frelich}, however, for diluted systems, such as atomic condensates, the difference is very small; it is small also for the superfluid helium. Therefore we will assume $\mathbf{E}_0 =E_0 \mathbf{e}$ to be the constant external electric field directed along the unit vector $\mathbf{e}$. The variable part appears when there is an electromagnetic wave.

The dipole moment is the sum of two terms
\begin{equation}
\mathbf{d}=\mathbf{d}_{0} +\mathbf{d}_{p} . 
\label{EQn_9_}
\end{equation}
Here $\mathbf{d}_{0}$ is the atom's intrinsic dipole moment, which in equilibrium is oriented along the external field. Its value $d_0$ is assumed to be constant. The second term is the dipole moment induced by the external field
\begin{equation}
\mathbf{d}_{p} 
	\equiv \mathbf{d}_{p} \left(t\right)
		=\int\limits_{-\infty }^{t}\alpha \left(t-t'\right) \mathbf{E}\left(t'\right)dt'
			=\int\limits_{0}^{\infty }\alpha \left(\tau \right) \mathbf{E}\left(t-\tau \right)d\tau , 
\label{EQn_10_}
\end{equation}
where $\alpha $ is the polarizability of the atom. The intrinsic dipole moment can, in general, change direction, so that $\mathbf{d}_0 =d_0 \mathbf{e}+\delta \mathbf{d}_0 (t)$, where the first term is the equilibrium moment and the second is the variable part. As the magnitude of $\mathbf{d}_0$ is constant, for small deviations from equilibrium $\delta \mathbf{d}_0 \cdot \mathbf{e}=0$. As will be seen below, due to this condition the fluctuations of the intrinsic dipole moment drop out of the equations for the macroscopic wave function. Therefore in this approximation for small oscillations we can assume the intrinsic dipole moment $\mathbf{d}_0$ to be constant: $\mathbf{d}_0=d_0 \mathbf{e}$.

The polarization part of the dipole moment is
\[\mathbf{d}_{p} \left(t\right)
	=\alpha _{0} \mathbf{E}_{0}+\delta \mathbf{d}_{p} \left(t\right),\]
where $\alpha _{0}$ is the static polarizability of the atom (see eq. \eqref{polarizability}). The first term here is the constant dipole moment induced by the constant external field, and the second part is the variable part, induced by the variable field
\begin{equation}
\delta \mathbf{d}_{p} \left(t\right)
	=\int\limits_{0}^{\infty }\alpha \left(\tau \right) \delta \mathbf{E}\left(t-\tau \right)d\tau . 
\label{EQn_11_}
\end{equation}

Thus the full dipole moment can be presented as the sum of equilibrium and variable parts
\[\mathbf{d}\left(t\right)=d_{s} \mathbf{e}+\delta \mathbf{d}\left(t\right),\qquad
	d_{s} =d_{0} +\alpha _{0} E_{0}\]
Hereafter we will drop the subscript $p$ for the non-stationary dipole moment in \eqref{EQn_11_}: $\delta \mathbf{d}_{p} \left(t\right)\equiv \delta \mathbf{d}\left(t\right)$. 

In stationary equilibrium state the phase of order parameter $\psi_0$ can be chosen real, then  \eqref{EQn_6_} implies that
\begin{equation}
\psi _{0}^{2} \equiv n_{0} =\frac{\mu +d_{s} E_{0} }{g} . 
\label{EQn_12_}
\end{equation}

Taking into account \eqref{EQn_12_}, the GP equation that incorporates relaxation and interaction with electromagnetic field in the dipole approximation takes the following form
\begin{equation}
\hbar \left(i-\gamma \right)\dot{\psi }=-\frac{\hbar ^{2} }{2m} \Delta \psi +g\left(\left|\psi \right|^{2} -n_{0} \right)\psi +\left(d_{s} E_{0} -\mathbf{E}\cdot \mathbf{d}\right)\psi . 
\label{EQn_13_}
\end{equation}

Polarization vector
\begin{equation} \label{EQn_14_}
\mathbf{P}=\mathbf{d}\left|\psi \right|^{2}  
\end{equation} 
is also the sum of two terms, the constant and variable parts $\mathbf{P}=\mathbf{P}_{s} +\delta \mathbf{P}\left(t\right)$. The constant part is comprised of the spontaneous polarization of atoms with dipole moments and the polarization due to constant external field
\begin{equation}
\mathbf{P}_{s} 
	=d_{0} n_{0} \mathbf{e}+\alpha _{0} n_{0} E_{0} \mathbf{e}=d_{s} n_{0} \mathbf{e}. 
\label{EQn_15_}
\end{equation}

The variable part is
\begin{equation}
\delta \mathbf{P}
	=\psi _{0} d_{s} \left(\delta \psi +\delta \psi ^{*} \right)\mathbf{e}
		+n_{0} \int\limits_{0}^{\infty }\alpha \left(\tau \right)
			\delta \mathbf{E}\left(t-\tau \right)d\tau .
\label{EQn_16_}
\end{equation}
Consequently, the full electric displacement $\mathbf{D}=\mathbf{E}+4\pi \mathbf{P}$ is composed of the constant part $\mathbf{D}_s$ and the electric displacement  $\delta \mathbf{D}$ induced by the variable field:
\begin{align}
\mathbf{D}=&\mathbf{D}_{s} +\delta \mathbf{D} ,
\label{EQn_17_}\\
&\mathbf{D}_{s} =4\pi \mathbf{P}_{s} +\mathbf{E}_{0} 
	=\varepsilon _{0} \mathbf{E}_{0} +4\pi d_{0} n_{0} \mathbf{e}, \nonumber \\
	& \delta \mathbf{D}=\delta \mathbf{E}+4\pi \delta \mathbf{P} \nonumber .
\end{align}
Here $\varepsilon _{0} =1+4\pi \alpha _{0} n_{0} $ is the static permittivity. Note that in the presence of intrinsic dipole moment the dipole-dipole interaction of atoms is composed of the short-range and long-range components. Equation \eqref{EQn_13_} only takes into account interaction with the external field and with the field of electromagnetic wave, but neglects the contribution of dipole-dipole interaction. The role of the latter is studied in section \ref{Sec7}.

\section{Dielectric permittivity of BEC of atoms with dipole moments}

The relation \eqref{EQn_16_} and equation \eqref{EQn_13_} allow one to find the permittivity of BEC atoms with dipole moments as the function of frequency and wave vector. In order to do this, we linearize equation \eqref{EQn_13_}, leaving the terms linear in $\delta \psi =\psi -\psi _{0} $ and the electric field variation $\delta\mathbf{E}$:
\begin{equation}
\hbar \left(i-\gamma \right)\delta \dot{\psi }
	=-\frac{\hbar ^{2} }{2m} \Delta \delta \psi 
		+gn_{0} \left(\delta \psi +\delta \psi ^{*} \right)
		-\left(E_{0} \delta \mathbf{d}+d_{s} \delta \mathbf{E}\right)\cdot \mathbf{e}\psi _{0} . 
\label{EQn_18_}
\end{equation}

Hereafter it is convenient to use real quantities 
\begin{equation}
\delta \Psi \equiv \delta \psi +\delta \psi^{*} ,\quad
 \delta \Phi \equiv i(\delta \psi -\delta \psi ^{*} ). 
\label{EQn_19_}
\end{equation}

If one defines the absolute value and phase of the macroscopic wave function $\psi=\rho e^{i\chi}$, assuming that in equilibrium $\rho_0 =\psi_0$ and $\chi_0 =0$, it turns out that the functions introduced in \eqref{EQn_19_} are related to the fluctuations of these quantities as
\begin{equation}
\delta\Psi =2\delta \rho,\qquad
\delta\Phi =2\psi_0 \delta\chi.
\label{Eq19a}
\end{equation}

Then \eqref{EQn_18_} gives the system of equations for the real functions \eqref{EQn_19_}:
\begin{align} 
\label{EQn_20_} 
&\hbar \delta \dot{\Phi }-\hbar \gamma \delta \dot{\Psi }
	=-\frac{\hbar ^{2} }{2m} \Delta \delta \Psi +2gn_{0} \delta \Psi 
	-2\psi _{0} \left(d_{s} \delta \mathbf{E}+E_{0} \delta \mathbf{d}\right)\cdot \mathbf{e},\\
&\hbar \delta \dot{\Psi }+\hbar \gamma \delta \dot{\Phi }
	=+\frac{\hbar ^{2} }{2m} \Delta \delta \Phi .
\end{align} 
This system describes in the linear approximation the dynamics of the superfluid system in a weak electric field. 

Let us assume that the variable field changes as
\begin{equation}
	\delta \mathbf{E}\left(\mathbf{r},t\right)
	=\mathbf{b}e^{iQ\left(\mathbf{r},t\right)},\qquad
Q(\omega ,\mathbf{k})\equiv \mathbf{kr}-\omega t .
\label{Q}
\end{equation}
Then the solutions of the system \eqref{EQn_20_} also take this form
\[\delta \Psi (\mathbf{r},t)=\Psi _{0} e^{iQ\left(\mathbf{r},t\right)},\qquad
	\delta \Phi (\mathbf{r},t)=\Phi _{0} e^{i Q(\mathbf{r},t)},\]
with
\begin{align}
\Psi _{0} &=-2\psi _{0} \frac{\left(\varepsilon _{k} -i\gamma \hbar \omega \right)}{D} \left[d_{s} +\alpha \left(\omega \right)E_{0} \right]\mathbf{b}\cdot \mathbf{e},\\
 \Phi _{0} &=-2\psi _{0} \frac{i\hbar \omega }{D} \left[d_{s} +\alpha \left(\omega \right)E_{0} \right]\mathbf{b}\cdot \mathbf{e}, \label{EQn_21_}
\end{align}
where
\begin{align}
D\equiv D\left(\omega ,\mathbf{k}\right)
	\equiv &\left(1+\gamma ^{2} \right)\left(\hbar \omega \right)^{2}
		 -\varepsilon _{k} \left(\varepsilon _{k} +2gn_{0} \right)
		 +2i\gamma \hbar \omega \left(\varepsilon _{k} +gn_{0} \right);
\label{EQn_22_}
\end{align}
and 
\begin{equation}
\varepsilon _{k} =\frac{\hbar ^{2} k^{2}}{2m}
\end{equation}
is the kinetic energy of the free atom.

The Fourier image of the atom's polarizability
\begin{equation}
	\label{polarizability}
	\alpha \left(\omega \right)
	\equiv \int\limits_{0}^{\infty }\alpha \left(\tau \right) e^{i\omega \tau } d\tau
\end{equation}
is often chosen in the model of damped oscillator in the form
\begin{equation}
\alpha \left(\omega \right)
	=\frac{N_{a} e^{2} }{m_{0} } \frac{1}{\omega _{0}^{2} -\omega ^{2} -i\nu \omega } , 
\label{EQn_23_}
\end{equation}
where $N_{a} $ is the number of electrons in the atom which give contribution to polarization, $\nu$ is a phenomenological damping coefficient, $m_0$ electron mass, $\omega_0$ the characteristic frequency of electronic oscillations. In this model the static polarizability of the atom is
\[\alpha _{0} \equiv \alpha (0)
	=\frac{N_{a} e^{2}}{m_0 \omega_0^2}.\]
Its real and imaginary parts \eqref{EQn_23_} $\alpha \left(\omega \right)=\alpha '\left(\omega \right)+i\alpha ''\left(\omega \right)$ are
\begin{align}
\alpha '\left(\omega \right)&
	=\alpha _{0} \frac{\omega _{0}^{2} \left(\omega _{0}^{2} -\omega ^{2} \right)}
	{\left(\omega _{0}^{2} -\omega ^{2} \right)^{2} +\nu ^{2} \omega ^{2} } ,\\
 \alpha ''\left(\omega \right)&
	=\alpha _{0} \frac{\nu \omega \omega _{0}^{2} }
		{\left(\omega _{0}^{2} -\omega ^{2} \right)^{2} +\nu ^{2} \omega ^{2} } .
\label{Eq23a}
\end{align}

Hereafter in this work we are mostly interested in the effects at low frequencies $\omega ^{2} \ll\omega _{0}^{2} $, assuming also that $\nu \ll\omega _{0} $. In this case, taking into account terms linear in $\nu $, we have
\begin{equation}
\alpha '\left(\omega \right)\approx \alpha _{0} \left(1+\frac{\omega ^{2} }{\omega _{0}^{2} } \right),\quad \quad \alpha ''\left(\omega \right)\approx \alpha _{0} \frac{\nu \omega }{\omega _{0}^{2} }. 
\label{Eq23b}
\end{equation}

Using \eqref{EQn_21_}, one can rewrite the time-dependent polarization vector in the form
\[\delta \mathbf{P}\left(r,t\right)=\mathbf{p}e^{iQ\left(\mathbf{r},t\right)},
	\qquad p_{i} =\kappa _{ij} \left(\omega,k \right)b_{j}.\]
Then the polarizability tensor $\kappa_{ij}$ of the BEC takes the form
\begin{align}
\kappa _{ij} \left(\omega,k \right)=&
	\kappa _{0} \left(\omega \right)\delta _{ij} 
		+\kappa _{\bot } \left(\omega,k \right)e_{i} e_{j} \label{EQn_24_};\\
&\kappa _{0} \left(\omega \right)
	=n_{0} \alpha \left(\omega \right),\\
&\kappa _{\bot } \left(\omega ,k\right)
	=\frac{2d_{s} n_{0} }{D\left(\omega ,k\right)} 
		\left[d_{s} +\alpha \left(\omega \right)E_{0} \right]
			\left(i\gamma \hbar \omega -\varepsilon _{k} \right). \label{kappabot}
\end{align}
The permittivity tensor 
\[\varepsilon _{ij} \left(\omega,k \right)=\delta _{ij} +4\pi \kappa _{ij} \left(\omega,k \right)\]
then can be presented as
\begin{align}
\varepsilon _{ij} \left(\omega,k \right)
	=&\left[1+4\pi n_{0} \alpha \left(\omega \right)\right]\delta _{ij} 
		+\varepsilon _{\bot } \left(\omega ,k\right)e_{i} e_{j} , 
\label{EQn_25_}\\
\label{EQn_26_} 
&\varepsilon _{\bot } \left(\omega ,k\right)
	=4\pi \kappa _{\bot } \left(\omega ,k\right)
		=\frac{8\pi d_{s} n_{0} }{D\left(\omega ,k\right)} 
			\left[d_{s} +\alpha \left(\omega \right)E_{0} \right]
				\left(i\gamma \hbar \omega -\varepsilon _{k} \right). 
\end{align} 
The isotropic part of it depends on frequency only, while the anisotropic part $\varepsilon_{\bot } \left(\omega ,k\right)$ is also a function of wave vector. Thus, the BEC of atoms with dipole moments is a medium with both temporal and spacial dispersion \cite{Agra}. 

We will consider propagation of electromagnetic waves of given frequency, assuming thus that $\omega$ is real. Due to dissipation the wave vector will then be complex $\mathbf{k}=\mathbf{k}'+i\mathbf{k}''$. In general, its real and imaginary parts can be directed differently, but we will limit our attention to homogeneous wave, in which $\mathbf{k}=\left(k'+ik''\right)\mathbf{s}$, where $\mathbf{s}$ is a real unit vector, which determines the direction of wave's propagation.

\section{Electromagnetic waves in BEC}
The Maxwell's equations
\begin{align}
&{\rm div}\delta \mathbf{D}=0,
	&&{\rm rot}\delta \mathbf{B}=\frac{1}{c} \frac{\partial \delta \mathbf{D}}{\partial t},
		\nonumber\\
& {\rm div}\delta \mathbf{B}=0 ,
	&&	{\rm rot}\delta \mathbf{E}=-\frac{1}{c} \frac{\partial \delta \mathbf{B}}{\partial t} ,
			 \label{EQn_27_} 
 \end{align} 
with the polarizability tensor of the form  \eqref{EQn_24_}, neglecting magnetic properties of the medium, give the following equation describing electromagnetic waves in BEC:
\begin{equation}
\left\{\Big[\frac{\omega ^{2} }{c^{2} } 
	\left(1+4\pi n_{0} \alpha \left(\omega \right)\right)-k^{2} \Big]\delta _{ij} 
		+k_{i} k_{j} 
		+4\pi \frac{\omega^2}{c^2} \kappa _{\bot }(\omega ,k)e_{i} e_{j} \right\}b_{j} =0. \label{EQn_28_}
\end{equation}

The dispersion laws of the propagating waves are obtained by equating the determinant of the system of homogeneous equations \eqref{EQn_28_} to zero. We align the $z$-axis along the external field, so that $\mathbf{e}=(0,0,1)$ and align the others so that the wave propagates in the $xz$ plane at angle $\theta$ to axis $z$. Then $\mathbf{k}=(k\sin\theta,0,k\cos\theta)$.

For a wave with electric field directed along the $y$-axis the dispersion law is given by relation
\begin{equation}
\frac{\omega ^{2} }{c^{2} } \varepsilon (\omega)=k^{2} , \label{EQn_29_}
	\qquad
	\varepsilon \left(\omega \right)
		\equiv \varepsilon '\left(\omega \right)+i\varepsilon ''\left(\omega \right)
			=1+4\pi n_{0} \alpha \left(\omega \right).
\end{equation}
At small frequencies $\omega ^{2} \ll\omega _{0}^{2} $ taking into account \eqref{Eq23b} the real and imaginary parts of permittivity are 
\begin{equation}
\varepsilon '\left(\omega \right)
	=\varepsilon _{0} +\frac{4\pi n_{0} \alpha _{0} \omega ^{2} }{\omega _{0}^{2} } ,
		\qquad
		\varepsilon ''\left(\omega \right)
		=\frac{4\pi n_{0} \alpha _{0} \nu \omega }{\omega _{0}^{2} },
\label{Eq29a}
\end{equation}
where $\varepsilon _{0} \equiv \varepsilon '\left(0\right)=1+4\pi n_{0} \alpha _{0} $ is the static permittivity. Equation \eqref{EQn_29_} does not contain the intrinsic dipole moment and can be analyzed in the same way as in the case of wave propagation in a medium with temporal dispersion \cite{Agra,Dav}. This wave does not couple to the condensate and the latter does not oscillate.

The case of more interest is when the electric field oscillates in the $xz$ plane. Then the dispersion law is given by
\begin{equation}
\left[k^{2} -\frac{\omega ^{2} }{c^{2} } \varepsilon \left(\omega \right)\right]\left[\varepsilon \left(\omega \right)+4\pi \kappa _{\bot } \left(\omega ,k\right)\cos ^{2} \theta \right]=4\pi \frac{\omega ^{2} }{c^{2} } \varepsilon \left(\omega \right)\kappa _{\bot } \left(\omega ,k\right)\sin ^{2} \theta , 
\label{EQn_30_}
\end{equation}
where $\kappa _{\bot }$ is defined in \eqref{kappabot}. The components of electric field's intensity are related by
\begin{equation} 
\label{EQn_31_} 
\Big[\frac{\omega ^{2} }{c^{2} } \varepsilon \left(\omega \right)
	-k^{2} \cos ^{2} \theta \Big]\; b_{x} +k^{2} \sin \theta \cos \theta \; b_{z} =0.
\end{equation} 

\subsection{Propagation of waves along the dipoles}
Let us at first consider the special case of wave propagating along the direction of the dipole moment ($\theta=0$). For the transversal wave, in which $b_x^2 +b_y^2 > 0$, the dispersion law is given by \eqref{EQn_29_}, while the condensate is not excited. For the longitudinal wave, in which $b_z \neq 0$, the dispersion law is given by relation
\begin{equation}
\varepsilon \left(\omega \right)+4\pi \kappa _{\bot } \left(\omega ,k\right)=0. \label{EQn_32_}
\end{equation}
In this wave the condensate oscillates together with the longitudinal component of the electric field.

Hereafter we will consider the external field to be weak, setting $E_0 \to 0$. Practically this means that external field is used only to distinguish the dipole moments' orientation. If one neglects dissipation ($\gamma\to 0$) and imaginary part of permittivity ($\varepsilon''\to 0$), \eqref{EQn_32_} gives the modified dispersion law for the excitations of the BEC
\begin{align}
\hbar ^{2} \omega ^{2} 
	=&E_{k}^{2} +\frac{8\pi d_{0}^{2} n_{0} }{\varepsilon(\omega)} \varepsilon _{k}, 
	\label{EQn_33_}\\
	&E_{k}^{2} =\varepsilon _{k} \left(\varepsilon _{k} +2gn_{0} \right) . \label{Ek}
\end{align}
Here $E_k$ is the dispersion law without taking into account the intrinsic dipole moment. 

Without dipole moment the relation \eqref{EQn_33_} gives the dispersion law of BEC excitations, which at low frequencies has the form
\[E_{k} =c_{B} \hbar k,\qquad c_{B} =\sqrt{gn_{0} /m},\]
where $c_B$ is the Bogoliubov sound velocity. Due to the existence of dipole moment the condensate oscillations are accompanied by oscillation of electric field, while the sound velocity is modified and becomes
\begin{equation}
\tilde{c}_B=\sqrt{\tilde{g}n_0 /m},\qquad
	\tilde{g}=g+\frac{4\pi d_{0}^{2} }{\varepsilon _{0} }.
 \label{EQn_34_}
\end{equation}

We used here that at small frequencies one can assume $\varepsilon'(\omega)\approx \varepsilon_{0} $. Thus taking into account electrical effects in a BEC of atoms with dipole moments leads to increase of the sound velocity, which is equivalent to the corresponding increase of the interaction constant $g$. Without dissipation the amplitudes of the electric field and condensate wave function oscillations are related according to \eqref{EQn_21_}:
\begin{equation}
\frac{\Psi _{0} }{\psi _{0} }
	 =-\frac{\varepsilon\left(\omega \right)}{4\pi d_{0} n_{0} } b_{z},
		\qquad
	\frac{\Phi _{0} }{\psi _{0} } 
		=-i\frac{\hbar \omega \varepsilon(\omega)}{4\pi d_{0} n_{0} \varepsilon _{k} } b_{z}. 
\label{EQn_35_}
\end{equation}

Now let us consider the effect of dissipation. In this case $k=k'+ik''$ and $\varepsilon _{k} =\varepsilon '_{k} +i\varepsilon ''_{k} $, where
\begin{equation}
\varepsilon '_{k} =\frac{\hbar ^{2} }{2m} \left(k'^{2} -k''^{2} \right),
	\qquad \varepsilon ''_{k} =\frac{\hbar ^{2} }{m} k'k''. \label{EQn_36_}
\end{equation}

Equation \eqref{EQn_32_}, taking into account terms linear in $\gamma $ and $\nu $, implies that the real and imaginary parts of energy as functions of energy are
\begin{align} 
{\varepsilon '}_{k}^{2} +2\lambda \left(\omega \right)\varepsilon '_{k} -\hbar ^{2} \omega ^{2} =0,\nonumber\\
{\varepsilon''}_{k} =\gamma \hbar \omega +8\pi d_{0}^{2} n_{0} \zeta \left(\omega \right), \label{EQn_37_} 
 \end{align} 
where
\[\lambda \left(\omega \right)
	\equiv n_{0} \left[g+\frac{4\pi d_{0}^{2} }{\varepsilon '\left(\omega \right)} \right],
\qquad
\zeta \left(\omega \right)
	\equiv \left[1+\left(\frac{\hbar \omega }{\varepsilon '_{k} } \right)^{2} \right]^{-1}
	 \frac{\varepsilon ''\left(\omega \right)}{\varepsilon '^{2} \left(\omega \right)} .\] 

Equation \eqref{EQn_37_}, with relations \eqref{EQn_36_}, allow one to find the real and imaginary parts of the wave vector as functions of frequency
\begin{equation}
k'^{2} =\frac{m}{\hbar ^{2} } 
	\Big[\sqrt{1+\left(A/B\right)^{2} } +1\Big],
		\qquad
		 k''^{2} =\frac{m}{\hbar ^{2} } \Big[\sqrt{1+\left(A/B\right)^{2} } -1\Big], 
\label{EQn_38_}
\end{equation}
where the following notation is introduced
\[A\equiv \lambda (\omega )
	\Big\{\sqrt{1+\big[\hbar \omega /\lambda (\omega )\big]^{2} } -1\Big\},
\qquad
 B\equiv \gamma \hbar \omega +8\pi d_{0}^{2} n_{0} \zeta \left(\omega \right).\] 

At low frequencies, such that $\omega ^{2} \ll\omega _{0}^{2}$, when the condition $\hbar \omega \ll\tilde{g}n_{0} $ also holds, one has
\begin{equation}
k'^{2} =\frac{\omega ^{2} }{2\tilde{c}_{B}^{2} } \left[\sqrt{1+\left(\frac{2\tilde{\gamma }n_{0} \tilde{g}}{\hbar \omega } \right)^{2} } +1\right],\quad \quad \quad k''^{2} =\frac{\omega ^{2} }{2\tilde{c}_{B}^{2} } \left[\sqrt{1+\left(\frac{2\tilde{\gamma }n_{0} \tilde{g}}{\hbar \omega } \right)^{2} } -1\right], 
\label{EQn_39_}
\end{equation}
where $\tilde{\gamma}$ is a new dissipative coefficient:
\begin{equation}
\tilde{\gamma }\equiv \gamma +\gamma _{1}, \qquad
\gamma _{1} \equiv \frac{8\pi d_{0}^{2} n_{0} }{\hbar \omega _{0} } 
	\frac{\left(\varepsilon _{0} -1\right)}{\varepsilon _{0}^{2} } \frac{\nu }{\omega _{0} }. 
\label{EQn_40_}
\end{equation}

Here $\gamma$ takes into account the dissipation due to relaxation of the condensate, while $\gamma_1$ appears due to imaginary part of permittivity. In order to estimate the value of $\gamma_1$, we take $d_{0} =el$, where $l\sim 10^{-8} {\rm cm}$, $n_{0} =10^{22}\,{\rm cm}^{-3} $, ${\omega _{0}/2\pi }\sim 10^{15} {\rm Hz}$, $\nu \sim 10^{8} {\rm Hz}$, $\left(\varepsilon _{0} -1\right)/ \varepsilon _{0}^{2} \approx 10^{-1} $. This yields $\gamma _{1} \sim 10^{-9} \doteq 10^{-8} $. As mentioned above, some estimates \cite{YM14} provide $\gamma \sim 10^{-6} \doteq 10^{-5} $, thus probably $\gamma _{1} \ll \gamma $. 

At low frequencies, such that $\hbar \omega \ll \gamma n_{0} \tilde{g}$, 
\begin{equation} \label{EQn_41_} 
k'^{2} \sim k''^{2} \sim 
	\tilde{\gamma }\frac{n_{0} \tilde{g}\omega }{\hbar \tilde{c}_{B}^{2} }  
\end{equation} 
and propagation of excitations has relaxational character.

The dispersion law of transverse waves \eqref{EQn_29_} does not contain the contribution of dipole moments and propagation of such waves can be considered in the same way as propagation of waves in a atomic gas with polarizability \eqref{EQn_23_} \cite{Agra,Dav}. 

\subsection{Propagation of waves orthogonal to dipoles}
In another special case of $\theta =\pi/2$, when the wave propagates along the $x$ axis, longitudinal waves with $b_x \neq 0$ are absent, while for the transversal wave with $b_z \neq 0$ the dispersion law is determined by equation
\begin{equation}
\left[\varepsilon _{k} 
	-\frac{\hbar ^{2} \omega ^{2} }{2mc^{2} } \varepsilon \left(\omega \right)\right]
		D\left(\omega ,k\right)
	=\eta \hbar ^{2} \omega ^{2} \left(i\gamma \hbar \omega -\varepsilon _{k} \right).
 \label{EQn_42_}
\end{equation}

Here we have introduced a dimensionless parameter
\begin{equation}
\eta \equiv \frac{4\pi d_{0}^{2} n_{0} }{mc^{2} }, 
\label{EQn_43_}
\end{equation}
which regulates the coupling of condensate to the field. For its estimate we take $n_{0} =10^{22}{\rm cm}^{-3} $, $m=10^{-23} {\rm g}$, $d_{0} =el$, where $l\sim 10^{-8} {\rm cm}$ as before, which gives $d_{0} \approx 5\cdot 10^{-18} {\rm cm}^{5/2} {\rm g}^{1/2} {\rm s}^{-1} $, which is close to the characteristic value for polar molecules. Then we have $\eta \sim 10^{-10} $. It is of interest to find the ratio of $\eta$ \eqref{EQn_43_} to polarizability: 
\begin{equation}
\frac{\eta }{\varepsilon _{0} -1} =N_{a}^{-1} \cdot \frac{m_{0} }{m} \cdot \left(\frac{l\omega _{0} }{c} \right)^{2} \sim 10^{-9}. 
\label{EQn_44_}
\end{equation}
Here we used the following values: $N_{a} \sim 1$, $m_{0} /m \sim 10^{-3} $, $\omega _{0} \sim 3\cdot 10^{15} {\rm Hz}$. Frequency $\omega _{0} $ corresponds to transition between levels with energy difference of the order of $10\; {\rm eV}$. 

Due to the smallness of parameter $\eta $, the solution of equation \eqref{EQn_42_} can be presented in the form of expansion by $\eta$
\begin{equation}
\varepsilon _{k} =\varepsilon _{k}^{(0)} +\eta \varepsilon _{k}^{(1)}.
\label{EQn_45_}
\end{equation}
In the same approximation, leaving only terms linear in dissipative coefficients, we get
\begin{align}
D\left(\omega ,k\right)=&D^{(0)} \left(\omega ,k\right)+\eta D^{(1)} (\omega ,k),\\
	\label{EQn_46_} 
	&D^{\left(0\right)} \left(\omega ,k\right)
		=\left(\hbar \omega \right)^{2} 
			-\varepsilon _{k}^{\left(0\right)} \left(\varepsilon _{k}^{\left(0\right)} 
			+2gn_{0} \right)+2i\gamma \hbar \omega \left(\varepsilon _{k}^{\left(0\right)}
			 +gn_{0} \right), \\
	&D^{\left(1\right)} \left(\omega ,k\right)
		=2\varepsilon _{k}^{\left(1\right)} \left[i\gamma \hbar \omega 
			-\left(\varepsilon _{k}^{\left(0\right)} +gn_{0} \right)\right].
\end{align}
In the zeroth order there are two modes, corresponding to independent propagation of field and condensate excitations. Existence of atom's dipole moments leads to coupling of these modes in the next approximation.

Let us consider first the solution, which describes in the zeroth order the propagation of electromagnetic field:
\begin{equation}
\varepsilon _{k}^{\left(0\right)} 
	=\frac{\hbar ^{2} \omega ^{2} }{2mc^{2} } \varepsilon \left(\omega \right). 
\label{EQn_47_}
\end{equation}
In this case the first order correction is
\begin{equation}
\varepsilon _{k}^{\left(1\right)} 
	=-\hbar ^{2} \omega ^{2} 
		\frac{\varepsilon _{k}^{(0)}-i\gamma \hbar \omega }
			{D^{\left(0\right)} \left(\omega ,k\right)}. 
\label{EQn_48_}
\end{equation}

At low frequencies, such that ${\varepsilon '}_{k}^{\left(0\right)} \ll gn_{0} $ or equivalently $\hbar ^{2} \omega ^{2} \ll m^{2} c^{2} c_{B}^{2} $, and neglecting both the quantities of the order of $(c_{B}/c )^{2} $ and dissipation, we have
\begin{equation}
\varepsilon '_{k} 
	=\frac{\hbar ^{2} \omega ^{2} }{2mc^{2} } \varepsilon _{0} \left(1-\eta \right),
 \label{EQn_49_}
\end{equation}
so that electromagnetic wave propagates with velocity
\begin{equation}
c_{*} =c\sqrt{\frac{1+\eta }{\varepsilon _{0} } }. 
\label{EQn_50_}
\end{equation}
The smallness of parameter $\eta $ ensures that for any reasonable values of parameters  $c_{*} <c$. 

Let us now consider the solution, which in the zeroth order describes excitations of the condensate 
\begin{equation}
\left(\hbar \omega \right)^{2} -\varepsilon _{k}^{\left(0\right)} \left(\varepsilon _{k}^{\left(0\right)} +2gn_{0} \right)+2i\gamma \hbar \omega \left(\varepsilon _{k}^{\left(0\right)} +gn_{0} \right)=0. 
\label{EQn_51_}
\end{equation}

Neglecting dissipation \eqref{EQn_51_} one obtains the Bogoliubov dispersion law for BEC \cite{Pit} $\hbar ^{2} \omega ^{2} ={\varepsilon '}_{k}^{\left(0\right)} \left({\varepsilon '}_{k}^{\left(0\right)} +2gn_{0} \right)$, so
\begin{equation}
{\varepsilon '}_{k}^{\left(0\right)} 
	=gn_{0} \left[\sqrt{1+\left(\frac{\hbar \omega }{gn_{0} } \right)^{2} } -1\right]. 
\label{EQn_52_}
\end{equation}

Damping is described by the imaginary part of excitations' energy 
${\varepsilon ''}_{k}^{\left(0\right)} =\gamma \hbar \omega $. The first order correction in this case is given by
\begin{equation}
\varepsilon _{k}^{\left(1\right)} =\frac{\hbar ^{2} \omega ^{2} \left(i\gamma \hbar \omega -\varepsilon _{k}^{\left(0\right)} \right)}{2\left(\varepsilon _{k}^{\left(0\right)} -\frac{\hbar ^{2} \omega ^{2} }{2mc^{2} } \varepsilon \left(\omega \right)\right)\left(i\gamma \hbar \omega -\varepsilon _{k}^{\left(0\right)} -gn_{0} \right)}. 
\label{EQn_53_}
\end{equation}

Without dissipation the correction to excitation's energy due to dipole moments is
\begin{equation}
{\varepsilon '}_{k}^{\left(1\right)} ={\varepsilon '}_{k}^{\left(0\right)} \frac{\hbar ^{2} \omega ^{2} }{2\left({\varepsilon '}_{k}^{\left(0\right)} +gn_{0} \right)\left[{\varepsilon '}_{k}^{\left(0\right)} -\frac{\hbar ^{2} \omega ^{2} }{2mc^{2} } {\varepsilon '}\left(\omega \right)\right]}. \label{EQn_54_}
\end{equation}

At small frequencies $k^{2} =\omega ^{2}/c_{B*}^{2}$, where omitting quantities of the order of $(c_{B}/c)^{2} $, one has
\begin{equation}
c_{B*}^{2} =c_{B}^{2} \left(1-\eta \right). 
\label{EQn_55_}
\end{equation}

The sound velocity of the condensate modified by dipole effects turns out to be less than the Bogoliubov velocity $c_{B} =\sqrt{gn_{0}/m} $. The amplitudes of field and condensate oscillations are related through
\begin{align} 
\frac{\Psi _{0} }{\psi _{0} } =\frac{2d_{0} mc_{B}^{2} }{\eta {\kern 1pt} \hbar ^{2} \omega ^{2} } \left[\sqrt{1+\left(\frac{\hbar \omega }{mc_{B}^{2} } \right)^{2} } -1-\varepsilon \left(\omega \right)\frac{\hbar ^{2} \omega ^{2} }{2m^{2} c^{2} c_{B}^{2} } \right]b_{z} ,\\
\frac{\Phi _{0} }{\psi _{0} } =i\frac{2d_{0} }{\eta {\kern 1pt} \hbar \omega } \left\{1-\frac{\varepsilon \left(\omega \right)}{2} \left(\frac{c_{B} }{c} \right)^{2} \left[\sqrt{1+\left(\frac{\hbar \omega }{mc_{B}^{2} } \right)^{2} } +1\right]\right\}b_{z} .
\label{EQn_56_} 
\end{align} 

At small frequencies $\hbar \omega \ll mc_{B}^{2} $
\begin{equation}
b_{z} =4\pi d_{0} n_{0} \left(\frac{c_{B} }{c} \right)^{2} \frac{\Psi _{0} }{\psi _{0} } =-i2\pi d_{0} n_{0} \frac{\hbar \omega }{mc^{2} } \frac{\Phi _{0} }{\psi _{0} } . 
\label{EQn_57_}
\end{equation}

\subsection{The general case}
When the wave propagates at arbitrary angle to the direction of the external field, both modes described by Eqs.  \eqref{EQn_30_}, \eqref{EQn_31_} become not purely longitudinal or transversal. Without external field and neglecting dissipation Eq. \eqref{EQn_30_} takes form
\begin{equation}
\left[\omega ^{2} -\frac{k^{2} c^{2} }{\varepsilon(\omega)} \right]
	\left(\omega ^{2} -\frac{E_{k}^{2} }{\hbar ^{2} } \right)
		=\frac{\eta}{\varepsilon(\omega)}
		 \left[\omega ^{2} -\frac{k^{2} c^{2} }{\varepsilon(\omega)} 
			\cos ^{2} \theta \right]c^{2} k^{2}, 
\label{EQn_58_}
\end{equation}
where $E_k$ is defined in \eqref{Ek}. 

In the zeroth order by $\eta$ and $(c_B /c)^2$ the dispersion law of the electromagnetic wave is given by
\begin{equation}
\omega ^{2} 
	=\frac{c^{2} k^{2} }{\varepsilon \left(\omega \right)} 
		\big(1+\eta \sin ^{2}\theta \big). 
\label{EQn_59_}
\end{equation}
The projections of field's intensity in the same approximation are related as
\begin{equation}
b_{z} =- (1+\eta)\tan\theta \; b_{x} . 
\label{EQn_60_}
\end{equation}

Propagation of the electromagnetic wave is accompanied by oscillation of the condensate
\begin{equation}
	\Psi _{0} =-2d_{0} \sqrt{n_{0} } 
		\frac{\varepsilon _{k} }{\hbar ^{2} \omega ^{2} -E_{k}^{2} } b_{z} ,
		\qquad
	 \Phi _{0} =-2id_{0} \sqrt{n_{0} } 
		 \frac{\hbar \omega }{\hbar ^{2} \omega ^{2} -E_{k}^{2} } b_{z}. 
	 \label{EQn_61_}
\end{equation}
In the long wavelength limit $\omega =c_{d} k$, where $c_d$ is the light velocity modified due to hybridization, which discarding terms of the order of $(c_b /c )^2$ is 
\begin{equation}
c_{d}^{2} =\frac{c^{2} }{\varepsilon _{0} } 
	\left[1+\eta \sin ^{2} \theta \right]. 
\label{EQn_62_}
\end{equation}

The correction to light velocity due to hybridization in BEC is always positive, however the resulting light velocity is still always less than $c$. Indeed, Eq. \eqref{EQn_62_} can be rewritten as
\begin{equation}
\left(\frac{c_{d} }{c} \right)^{2} 
	=1-\frac{4\pi n_{0} e^{2} }{m_{0} \omega _{0}^{2} } 
		\left[N_{a}
			-\frac{m_{0} }{m} \left(\frac{l\omega _{0} }{c} \right)^{2} \right]. 
\label{EQn_63_}
\end{equation}

For any reasonable values of parameters the expression in the brackets is positive, therefore light velocity is always less than in vacuum, as should be.

Let us consider the waves in condensate taking anto account hybridization. In the first order by $\eta$ and $(c_B /c)^2$ the dispersion law is given by
\begin{equation}
\omega ^{2} 
	=\frac{E_{k}^{2} }{\hbar ^{2} } +\eta 
	\frac{c^{2} k^{2}}{\varepsilon(\omega)}\cos^2 \theta. 
\label{EQn_64_}
\end{equation}
Projections of field intensity are related through
\begin{equation} 
\label{EQn_65_} 
b_{x} =\tan\theta\; b_{z}  .
\end{equation} 

The oscillations of the condensate are related to oscillations of electromagnetic field: 
\begin{equation}
\Psi _{0} =-2\sqrt{n_{0} } d_{0} 
	\frac{\varepsilon _{k} }{\hbar ^{2} \omega ^{2} -E_{k}^{2}} b_{z} ,
		\qquad
\Phi _{0} =-2i\sqrt{n_{0} } d_{0} 
	\frac{\hbar \omega }{\hbar ^{2} \omega ^{2} -E_{k}^{2}} b_{z}. 
			\label{EQn_66_}
\end{equation}

In the long wavelength limit $\omega =c_{Bd} k$, where $c_{Bd}$ is the condensate sound velocity, modified due to hybridization through dipole moment, which, neglecting terms of the order of $(c_{B}/c)^{2} $, is given by
\begin{equation}
c_{Bd}^{2} =c_{B}^{2} \left[1
	+\frac{\eta}{\varepsilon_0} \Big(\frac{c_B}{c}\Big)^{2} \cos^2 \theta \right]. 
\label{EQn_67_}
\end{equation}

Thus, the speed of wave's propagation in the condensate is increased due to hybridization with the electromagnetic wave through the dipole moment. That said, the angle dependent correction can in general be not small, in spite of the smallness of parameter $\eta$ \eqref{EQn_43_}. The formula \eqref{EQn_67_} only holds, however, if $\eta (c_B /c)^2 \cos^2\theta \ll 1$.

\section{Hybridization of electromagnetic waves and BEC oscillations in the presence of energy gap}
We have considered the influence of spontaneous polarization on the propagation of electromagnetic waves and the excitation spectrum of the condensate for the case when excitations are sound-like, with linear spectrum at low frequencies. In the experiment \cite{Ryb09} a resonance absorption of microwave radiation was discovered in superfluid helium at frequency close to 180 GHz, which was interpreted in \cite{YM14} as an indication of the existence, along with the sound excitations, of elementary excitations with an energy gap. Let us note that a branch of excitations with energy gap can, for example, exist in the condensate of atoms and their diatomic bound states \cite{PPP14}. In the presence of excitations with a gap their dispersion curve intersects with the one of electromagnetic waves, which leads to the phenomena of spectra hybridization and resonant absorption of radiation \cite{YM14}. The equation \eqref{EQn_58_} in this case remains valid, if $ E_ {k} $ is interpreted as the new dispersion law with the gap. For simplicity, we will neglect the dependence of dispersion law on wave vector, assuming $ E_ {k}=\Delta $. By neglecting also the dispersion of the permittivity we get $\varepsilon \left (\omega \right) \approx \varepsilon _ {0} $. Then equation \eqref{EQn_58_}, which determins the dispersion laws, becomes
\begin{equation}
	\left(\omega ^{2} -\frac{c^{2} k^{2} }{\varepsilon _{0} } \right)\left(\omega ^{2} -\omega _{*}^{2} \right)=\frac{\eta }{\varepsilon _{0} } \left(\omega ^{2} -\frac{c^{2} k^{2} }{\varepsilon _{0} } \cos ^{2} \theta \right)c^{2} k^{2} ,
	 \label{EQn_68_}
\end{equation}
where $\omega _{*}^{2} =\Delta ^{2}/\hbar ^{2}$. 

In the case of propagation in the direction of polarization vector ($\theta=0$), the transverse wave's dispersion law is given by Eq. \eqref{EQn_29_}, while the one for the longitudinal wave is modified and becomes
\begin{equation}
\omega ^{2} =\omega _{*}^{2} 
	+\frac{\eta }{\varepsilon _{0} } c^{2} k^{2} . \label{EQn_69_}
\end{equation}
The amplitudes of oscillations of the condensate and the electric field are related as
\begin{equation}
\Psi _{0} =\frac{\Delta }{i\varepsilon _{k} } \Phi _{0} 
	=-\frac{\varepsilon _{0} }{4\pi d_{0} \sqrt{n_{0} } } b_{z} . 
	\label{EQn_70_}
\end{equation}

In order to analyze the general case $\theta\neq 0$ it is convenient to rewrite equation \eqref{EQn_68_} in dimensionless form:
\begin{equation}
	\left(\tilde{\omega }^{2} -\tilde{k}^{2} \right)\left(\tilde{\omega }^{2} -1\right)
		=\eta \left(\tilde{\omega }^{2} -\tilde{k}^{2} \cos ^{2} \theta \right)\tilde{k}^{2} ,
 \label{EQn_71_}
 \end{equation}
where we have introduced the dimensionless frequency and wave vector
\begin{equation}
	\tilde{\omega }\equiv \frac{\omega }{\omega _{*} } ,
	\qquad \tilde{k}\equiv \frac{k}{k_{r} } ;
\label{EQn_72_}
\end{equation}
here $k_{r}^{2} =\varepsilon _{0} \omega _{*}^{2}/c^{2}$ is the squared wave vector corresponding to the intersection of the two unperturbed dispersion curves. Solution of quadratic equation \eqref{EQn_71_} gives us two branches
\begin{equation}
\tilde{\omega }_{\pm }^{2} =\frac{1}{2} \left\{\left(1+\sigma \tilde{k}^{2} \right)\pm \sqrt{\left[1+\left(\sigma -2\right)\tilde{k}^{2} \right]^{2} +4\left(\sigma -1\right)\tilde{k}^{4} \sin ^{2} \theta } \right\},
\label{EQn_73_}
\end{equation}
where $\sigma =1+\eta $. The branch $\tilde{\omega }_{+}^{2} $ always lies above the branch $\tilde{\omega }_{-}^{2} $ because
\begin{equation}
\tilde{\omega }_{+}^{2} -\tilde{\omega }_{-}^{2} =\sqrt{\left[1+\left(\sigma -2\right)\tilde{k}^{2} \right]^{2} +4\left(\sigma -1\right)\tilde{k}^{4} \sin ^{2} \theta } , \label{EQn_74_}
\end{equation}
At small wave vectors $\tilde{k}\ll 1$ Eq. \eqref{EQn_73_} implies the relations
\begin{equation}
\tilde{\omega }_{+}^{2} \approx 1+\left(\sigma -1\right)\tilde{k}^{2} ,\quad \quad \quad \tilde{\omega }_{-}^{2} \approx \tilde{k}^{2}, 
\label{EQn_75_}
\end{equation}
so in the long wavelength limit the dependence on the angle disappears. In the opposite limiting case of large wave vectors $\tilde{k}\gg 1$ we find:
\begin{equation}
\tilde{\omega }_{+}^{2} \approx \left[1+\left(\sigma -1\right)\sin ^{2} \theta \right]\cdot \tilde{k}^{2} ,\quad \quad \quad \tilde{\omega }_{-}^{2} \approx \left(\sigma -1\right)\cos ^{2} \theta \cdot \tilde{k}^{2}. 
\label{EQn_76_}
\end{equation}

In the case of wave propagating normal to the direction of polarization vector ($\theta =\pi /2$), the frequency of the low frequency mode tends to constant value $\tilde{\omega }_{-}^{2} \approx \sigma ^{-1} $ when $\tilde{k}\gg 1$. 

Let us consider the two branches in the vicinity of $\tilde{k}=1$, where in the absence of hybridization they would intersect. The hybridization due to dipole moment leads to the difference in frequencies at $\tilde{k}=1$:
\begin{equation}
\tilde{\omega }_{\pm }^{2} =1+\frac{1}{2} \left[\eta \pm \sqrt{\eta ^{2} +4\eta \sin ^{2} \theta } \right], 
\label{EQn_77_}
\end{equation}
so that
\begin{equation}
\tilde{\omega }_{+}^{2} -\tilde{\omega }_{-}^{2} =\sqrt{\eta ^{2} +4\eta \sin ^{2} \theta }.
\label{EQn_78_}
\end{equation}

As implied by \eqref{EQn_78_}, the repulsion between the branches increases with $\theta$, is the smallest at $\theta=0$ and rises to $\tilde{\omega }_{+}^{2} -\tilde{\omega }_{-}^{2} =2\sqrt{\eta } $ at $\theta=\pi/2$. Using the same estimation as above \eqref{EQn_43_} and taking $\eta =10^{-10} $, for the frequency difference $\Delta \omega =\omega _{+} -\omega _{-} $ we obtain
\begin{equation}
	\frac{\Delta \omega }{\omega _{*} } \sim 10^{-5}. 
\label{EQn_79_}
\end{equation}
At frequencies $\omega_{*}/2\pi =180\, {\rm GHz}\sim 10^{12}\, {\rm Hz}$, corresponding to the observed resonance absorption in superfluid helium \cite{Ryb09}, we get $\Delta \omega\sim 10^{7} {\rm Hz}$. 

The amplitudes of the condensate's oscillations arerelated to the electric field as
\begin{equation} 
\label{EQn_80_} 
\frac{\Psi _{0} }{\sqrt{n_{0} } } 
	=-\frac{d_{0} \varepsilon _{0} }{mc^{2} } \frac{\tilde{k}^{2} }{\left(\tilde{\omega }^{2} -1\right)} b_{z} ,\quad \quad \frac{\Phi _{0} }{\sqrt{n_{0} } } =-i\frac{2d_{0} }{\hbar \omega _{*} } \frac{\tilde{\omega }}{\left(\tilde{\omega }^{2} -1\right)} b_{z} 
\end{equation} 
and have sharp maxima at $\tilde{\omega}=\tilde{\omega} _{\pm}$.

\section{Effects of dipole-dipole interaction}
\label{Sec7}
Thus far we have neglected the long-range dipole-dipole interaction, and in previous sections the interaction between atoms was considered point-like. Only interaction of the dipole moment with the external field and with the field of the electromagnetic wave was taken into account. It was assumed that this approximation works for the superfluid helium, where the atom's dipole moment is small. Otherwise, when the dipole moment is large enough, the effects due to dipole-dipole interaction can be essential. In this section we consider the effect of dipole-dipole interaction on the propagation of acoustic and electromagnetic waves in the BEC. In the non-local GP equation
\begin{equation}
\hbar \left(i-\gamma \right)\dot{\psi }
	=-\frac{\hbar ^{2} }{2m} \Delta \psi +\left[U(\mathbf{r},t)-\mu \right]\psi 
		+\psi \int U (\mathbf{r},\mathbf{r}')\left|\psi (\mathbf{r}',t)\right|d\mathbf{r}', 
			\label{EQn_81_}
\end{equation}
the potential energy of interaction between atoms
\begin{equation}
	U\left(\mathbf{r},\mathbf{r}'\right)
		=g\delta (\mathbf{r}-\mathbf{r}')+U_{D}(\mathbf{r},\mathbf{r}'), 
\label{EQn_82_}
\end{equation}
acquires the new long-range interaction term
\begin{equation}
	U_{D} (\mathbf{r},\mathbf{r}')
		\equiv \frac{d(\mathbf{r})\cdot d(\mathbf{r}')
			-3\big(\mathbf{x}\cdot \mathbf{d}(\mathbf{r})\big)
				\big(\mathbf{x}\cdot \mathbf{d}(\mathbf{r}')\big)}{R^{3} }, 
\label{EQn_83_}
\end{equation}
where $\mathbf{R}=\mathbf{r}'-\mathbf{r}$ and $\mathbf{x}\equiv \mathbf{R}/R$. It can be checked that introduction of the dipole-dipole interaction does not affect the equilibrium density. Expanding an atom's dipole moment $d(\mathbf{r})=d_{s} \mathbf{e}+\delta \mathbf{d}(\mathbf{r})$, we find the potential energy in the linear approximation by the dipole moment fluctuations $\delta \mathbf{d}$:
\begin{align} 
	\label{EQn_84n_} 
	U_{D}(\mathbf{r},\mathbf{r}')
		&= U_{D}^{(0)} +U_{D}^{(1)} (\mathbf{r},\mathbf{r}')+\ldots\\
	U_{D}^{\left(0\right)} 
		&=\frac{d_{s}^{2} }{R^{3} } \left[1-3(\mathbf{x}\cdot \mathbf{e})^{2} \right],
			\label{EQn_84a_} \\
	U_{D}^{\left(1\right)} 
		&=\frac{d_{s}^{2} }{R^{3} } 
			\left[\mathbf{e}\cdot \delta \mathbf{d}(\mathbf{r})
				+\mathbf{e}\cdot \delta \mathbf{d}(\mathbf{r}')\right]
			-3\frac{d_{s}^{2} }{R^{3} } (\mathbf{e}\cdot \mathbf{x})
			\left[\mathbf{x}\cdot \delta \mathbf{d}(\mathbf{r})
				+\mathbf{x}\cdot \delta \mathbf{d}(\mathbf{r}')\right].
				\label{EQn_84b_} 
\end{align} 

Then the linearized system of equations for functions \eqref{EQn_19_} takes the form
\begin{align} 
\label{EQn_85_} 
\hbar \delta \dot{\Phi }-\hbar \gamma \delta \dot{\Psi }
	&=-\frac{\hbar ^{2} }{2m} \Delta \delta \Psi +2gn_{0} \delta \Psi 
		-2\psi _{0} (d_{s} \mathbf{e}\cdot \delta \mathbf{E}
			+E_{0} \mathbf{e}\cdot \delta \mathbf{d})+2\delta J_{D} ,\\
\hbar \delta \dot{\Psi }+\hbar \gamma \delta \dot{\Phi }
	&=\frac{\hbar ^{2} }{2m} \Delta \delta \Phi ,
\end{align} 
where
\begin{align} 
\delta J_{D} &=n_{0} \psi _{0} d_{s} \Big\{
	\int \frac{\mathbf{e}\cdot \delta \mathbf{d}(\mathbf{r})
					+\mathbf{e}\cdot \delta \mathbf{d}(\mathbf{r}')}{R^{3} }d\mathbf{r}'
		-3\int \frac{\mathbf{x}\cdot \delta \mathbf{d}(\mathbf{r})
					+\mathbf{x}\cdot \delta \mathbf{d}(\mathbf{r}')}{R^{3} }
				(\mathbf{e}\cdot \mathbf{x}) d\mathbf{r}'  \Big\}+ \notag\\ 
		&\qquad +n_{0} d_{s}^{2} 
		\int \frac{1-3(\mathbf{e}\cdot \mathbf{x})^{2}}{R^{3}}
				\delta \Psi (\mathbf{r}')d\mathbf{r}'. \label{EQn_86_} 
\end{align} 

Assuming the fluctuations of all variable quantities are plane waves, we have 
\[\delta \mathbf{d}(\mathbf{r}')=\delta \mathbf{d}(\mathbf{r})e^{-i\mathbf{k}(\mathbf{r}-\mathbf{r}')},\qquad 
\delta \Psi (\mathbf{r}')=\delta\Psi (\mathbf{r})e^{-i\mathbf{k}(\mathbf{r}-\mathbf{r}')}. \]
In this case
\begin{equation}
\delta J_{D} 
	=-\frac{4\pi d_{s} n_{0} }{3} \left(\delta _{ij} -3s_{i} s_{j} \right)
	\left[\psi _{0} e_{i} \delta d_{j} (\mathbf{r})
			+d_{s} e_{i} e_{j} \delta \Psi (\mathbf{r})\right]. 
\label{EQn_87_}
\end{equation}

Plugging in also
\[\delta\Psi (\mathbf{r})=\Psi _{0} e^{iQ(\mathbf{r},t)},\qquad 
\delta\Phi (\mathbf{r},t)=\Phi _{0} e^{iQ(\mathbf{r},t)},\qquad
\delta \mathbf{E}(\mathbf{r},t)=\mathbf{b}e^{iQ(\mathbf{r},t)},\]
where $Q(\omega,\mathbf{k})\equiv \mathbf{kr}-\omega t$ as introduced in \eqref{Q}, we arrive to the linear system
\begin{align}
 \label{EQn_88a_} 
-i\hbar \omega \Phi _{0} -(\varepsilon_{k} +2gn_{0} -u-i\hbar \omega \gamma )\Psi _{0} 
	&=-2\psi _{0}[d_{s} +\alpha (\omega) E_{0}]\mathbf{e}\cdot \mathbf{b}
		+\mathbf{L}\cdot \mathbf{b}, \\ 
 \label{EQn_88b_} 
(\varepsilon_{k}-i\hbar \omega \gamma)\Phi _{0} -i\hbar \omega \Psi _{0} 
	&=0,
\end{align} 
where quantities
\begin{align} 
\label{EQn_89_} 
u\equiv 
	&\frac{8\pi d_{s}^{2} n_{0} }{3} \big[1-3(\mathbf{e}\cdot \mathbf{s})^{2} \big], \\ 
L\equiv 
	&-2\psi _{0} \frac{4\pi d_{s} n_{0} }{3} \alpha (\omega) \mathbf{e}
		+8\pi \psi _{0} n_{0} d_{s} \alpha (\omega)(\mathbf{e}\cdot \mathbf{s})\mathbf{s}
\end{align} 
appear due to the dipole-dipole interaction. Let us estimate them in the absence of external field and determine the conditions when the effects of dipole-dipole interaction can be neglected. The summand $u$ can be discarded if $u\ll gn_0$, which is equivalent to $d_0^2 n_0 \ll mc_B^2$. For characteristic values $n_0 \sim 10^{21}{\rm cm}^{-3}$, $m\sim 10^{-23}{\rm g}$, $c_B \sim 10^{4}{\rm cm/s}$ this condition holds when the particle's dipole moment is much less than the characteristic dipole moment of the polar molecule. As $L\sim (\varepsilon-1) \psi_0 d_0$, this quantity can be neglected in \eqref{EQn_88a_} as long as the permittivity is close to unity. Note that for superfluid helium $\varepsilon \sim 1.06$. So, if the dipole moment is small and permittivity is close to unity, the effect of dipole-dipole interaction  is inessential, as was assumed in the previous sections.

The solution of \eqref{EQn_88a_}--\eqref{EQn_88b_} is
\begin{align} 
\label{EQn_90_}
\Psi _{0} &=\frac{\varepsilon _{k} -i\gamma \hbar \omega }{D}
	 \Big\{-2\psi _{0} \left[d_{s} +\alpha(\omega) E_{0} \right]\mathbf{b}\cdot \mathbf{e}
	 	+\mathbf{b}\cdot \mathbf{L}\Big\},\\
\Phi _{0} &=\frac{i\hbar \omega }{D} 
	\Big\{-2\psi _{0} \left[d_{s} +\alpha(\omega) E_{0} \right]	\mathbf{b}\cdot \mathbf{e}
	 	+\mathbf{b}\cdot \mathbf{L}\Big\},
\end{align} 
where
\begin{equation}
D\equiv D(\omega,\mathbf{k})
	\equiv (\hbar \omega )^{2} -\left(\varepsilon_{k}-i\hbar \omega \gamma\right)
		\left(\varepsilon _{k} +2gn_{0} -u-i\hbar \omega \gamma\right). 
\label{EQn_91_}
\end{equation}

Taking into account the relation \eqref{EQn_90_} and expression for the polarization amplitude  \eqref{EQn_16_} $\mathbf{p}=\psi _{0} d_{s} \Psi _{0} \mathbf{e}+n_{0} \alpha (\omega)\mathbf{b}$, we find the BEC polarizability tensor with the dipole-dipole interaction:
\begin{equation}
\kappa _{ij}(\omega ,\mathbf{k})
	=\kappa _{0} (\omega )\delta _{ij} 
	+\kappa _{\bot } (\omega ,k)e_{i} e_{j} 
	+\kappa _{d}(\omega ,k)(\mathbf{e}\cdot \mathbf{s})e_{i} s_{j}, 
\label{EQn_92_}
\end{equation}
where
\begin{align}
\label{EQn_93_} 
&\kappa _{0}(\omega)=n_{0} \alpha (\omega), \\
&\kappa _{\bot } (\omega ,k)
	=-2n_{0} d_{s} \frac{\varepsilon _{k} -i\hbar \omega \gamma }{D}
		 \Big[d_{s} +\alpha (\omega) E_{0} +\frac{4\pi d_{s} n_{0} }{3} \alpha(\omega)\Big], 	
		 	\\
&\kappa_{d} (\omega ,k)
	=8\pi d_{s}^{2} n_{0}^{2} \alpha \left(\omega \right)
		\frac{\varepsilon _{k} -i\hbar \omega \gamma }{D} . 
\end{align} 

Thus the permittivity tensor $\varepsilon _{ij} \left(\omega ,\mathbf{k}\right)=\delta _{ij} +4\pi \kappa _{ij}(\omega,\mathbf{k})$ can be presented in the form
\begin{equation}
\varepsilon _{ij} \left(\omega ,\mathbf{k}\right)
	=\varepsilon _{ij}^{s} \left(\omega ,\mathbf{k}\right)
		+\varepsilon _{ij}^{a} \left(\omega ,\mathbf{k}\right), \label{EQn_94_}
\end{equation}
where we distinguish the tensor's symmetric and anti-symmetric parts
\begin{align} \label{EQn_95_} 
\varepsilon _{ij}^{s} \left(\omega ,\mathbf{k}\right)&
	=\varepsilon _{ji}^{s} \left(\omega ,\mathbf{k}\right)= 
	\notag\\
	 &=\left[1+4\pi \kappa _{0} \left(\omega \right)\right]\delta _{ij} 
	 	+4\pi \kappa _{\bot } \left(\omega ,k\right)e_{i} e_{j} 
	 	+2\pi \kappa _{d}(\omega ,k)(\mathbf{e}\cdot \mathbf{s})(e_{i} s_{j} +e_{j} s_{i}), \\ 
\varepsilon _{ij}^{a} \left(\omega ,\mathbf{k}\right)&
	=-\varepsilon _{ji}^{a} \left(\omega ,\mathbf{k}\right)=
	\notag \\
	&=2\pi \kappa _{d}(\omega ,k)(\mathbf{e}\cdot \mathbf{s})(e_{i} s_{j} -e_{j} s_{i}).
\end{align} 

The dipole-dipole interaction leads to appearance of the antisymmetric part of the polarizability tensor.  Let us note, that in general the permittivity tensor does not have to be either symmetric or Hermitian \cite{Agra}. The proof of its symmetry, based on the symmetry of kinetic coefficients \cite{LLel}, is invalid in this case. An antisymmetric addition to the Hermitian permittivity tensor, for example in the magnetic field, makes the medium optically active or gyrotropic  \cite{LLel}. Phase velocities of waves with the right and left polarizations in a gyrotropic medium are different, which leads to rotation of the polarization plane of a linearly polarized wave. In our case in absence of dissipation the permittivity tensor turns out to be real and asymmetric, which, however, as shown below, does not lead to the effect of the polarization plane's rotation.

Propagation of electromagnetic waves in the medium is described by the following system of linear equations
\begin{equation}
\left[k^{2} \left(\delta _{ij} -s_{i} s_{j} \right)
	-\frac{\omega ^{2} }{c^{2} }\varepsilon _{ij}\right]b_{j} =0. 
\label{EQn_96_}
\end{equation}

Let us align $z$-axis along the vector $\mathbf{e}$, and the coordinate system so that $\mathbf{s}$ lies in the $xz$ plane $\mathbf{s}=(\sin \theta ,0,\cos \theta )$. Then the permittivity tensor has the following components:
\begin{align}
\label{EQn_97A_} 
\varepsilon_{ij}^{s}&=\varepsilon\delta_{ij}
	+4\pi \big[\kappa_\bot 
		+\kappa_d \cos^2 \theta \big]\delta_i^z \delta_j^z
		+2\pi \kappa_d \cos\theta \sin\theta\; \delta_i^z \delta_j^x ;\\
		&\varepsilon \equiv \varepsilon (\omega)=1+4\pi \kappa_0 (\omega),\quad
		\kappa_\bot\equiv \kappa_\bot (\omega, k),\quad
		\kappa_d\equiv \kappa_d (\omega, k).
\end{align}
We see that at $\theta=0$ and $\theta=\pi/2$ the antisymmetric part turns to zero.

In our case it turns out to be convenient to reformulate the problem in terms of induction $\delta \mathbf{D}(\mathbf{r},t)=\mathbf{d}e^{iQ(\mathbf{r},t)}$, such that $d_{i} =\varepsilon _{ij} b_{j} $. The inverse tensor of dielectric permittivity $\eta _{ij} \equiv \varepsilon _{ij}^{-1}$, such that $\eta _{ik} \varepsilon _{kj} =\delta _{ij} $, gives the inverse transformation $b_{i} =\eta _{ij} d_{j} $. It has the form
\begin{equation}
	\eta _{ij} \equiv \eta _{ij}(\omega ,\mathbf{k})
		=\frac{1}{\varepsilon } \Big[\delta _{ij} -4\pi 
			\frac{\kappa _{\bot } e_{i} e_{j} +\kappa _{d} (\mathbf{es})e_{i} s_{j} }
				{\varepsilon +4\pi \kappa _{\bot } +4\pi \kappa _{d} (\mathbf{es})^{2} } \Big]. 
\label{EQn_98_}
\end{equation}
The tensor $\eta_{ij}$ is also not symmetric and has the following non-zero components:
\begin{align}
 \label{EQn_99a_} 
 \eta _{xx}& =\eta _{yy} =\frac{1}{\varepsilon } ,\\
 \eta _{zz}& =\frac{1}{\varepsilon +4\pi \kappa _{\bot } 
 									+4\pi \kappa _{d} \cos ^{2} \theta }, \label{EQn_99b_} \\
 \eta _{zx} &=-\frac{4\pi \kappa _{d} \sin \theta \cos \theta }{\varepsilon \left(\varepsilon +4\pi \kappa _{\bot } +4\pi \kappa _{d} \cos ^{2} \theta \right)} \label{EQn_99c_} .
 \end{align} 

The components of the induction amplitude obey the system of equation, which follows from \eqref{EQn_96_}:
\begin{equation}
\left(\frac{\omega ^{2} }{c^{2} } \delta _{ik} -k^{2} \vartheta _{ik} \right)d_{k} =0, \label{EQn_100_}
\end{equation}
where $\vartheta _{ik} \equiv \left(\delta _{ij} -s_{i} s_{j} \right)\eta _{jk} $. As $s_{i} \vartheta _{ik} =0$, Eq. \eqref{EQn_100_} implies that $(\mathbf{sd})=0$, thus the induction vector is normal to the direction of the wave's propagation. The non-zero components of the $\vartheta _{ik} $ tensor are
\begin{align}
 \label{EQn_101a_} 
 \vartheta _{xx} &
 	=\frac{\cos ^{2} \theta \left(\varepsilon +4\pi \kappa _{\bot } +4\pi \kappa _{d} \right)}{\varepsilon \left(\varepsilon +4\pi \kappa _{\bot } +4\pi \kappa _{d} \cos ^{2} \theta \right)},\\
 	\vartheta _{yy} &
 	=\frac{1}{\varepsilon } , \label{EQn_101b_} \\
 	\vartheta _{zz} &=\frac{\sin ^{2} \theta }{\left(\varepsilon +4\pi \kappa _{\bot } +4\pi \kappa _{d} \cos ^{2} \theta \right)} , \label{EQn_101c_} \\
 	\vartheta _{xz} &
 		=-\frac{\sin \theta \cos \theta }{\left(\varepsilon +4\pi \kappa _{\bot } +4\pi \kappa _{d} \cos ^{2} \theta \right)} , \label{EQn_101e_} \\
 	 \vartheta _{zx} &=-\frac{\sin \theta \cos \theta \left(\varepsilon +4\pi \kappa _{\bot } +4\pi \kappa _{d} \right)}{\varepsilon \left(\varepsilon +4\pi \kappa _{\bot } +4\pi \kappa _{d} \cos ^{2} \theta \right)} . \label{EQn_101f_} 
\end{align} 

The condition
\begin{equation}
\det \left[k^{2} \vartheta _{ik} -\frac{\omega ^{2} }{c^{2} } \delta _{ik} \right]=0, \label{EQn_102_}
\end{equation}
implies that the dispersion law for the wave with $d_y \neq 0$ is
\begin{equation}
	\frac{\omega ^{2} }{c^{2} } \varepsilon =k^{2}, 
\label{EQn_103_}
\end{equation}
while the condensate's oscillations in it are absent. 

The dispersion relation for the wave with the induction vector in the $xz$ plane is given by equation 
\begin{equation}
\left(\frac{\omega ^{2} }{c^{2} k^{2} } -\vartheta _{xx} \right)\left(\frac{\omega ^{2} }{c^{2} k^{2} } -\vartheta _{zz} \right)-\vartheta _{xz} \vartheta _{zx} =0. 
\label{EQn_104_}
\end{equation}

One can check that Eqs. \eqref{EQn_101a_}--\eqref{EQn_101f_} imply $\vartheta _{xx} \vartheta _{zz} =\vartheta _{xz} \vartheta _{zx} $, and that means that Eq. \eqref{EQn_104_} gives either the trivial solution $\omega ^{2} =0$ or the dispersion law determined by
\begin{equation}
\frac{\omega ^{2} }{c^{2} k^{2} } \varepsilon =1-\frac{4\pi \kappa _{\bot } \sin ^{2} \theta }{\varepsilon +4\pi \kappa _{\bot } +4\pi \kappa _{d} \cos ^{2} \theta }. 
\label{EQn_105_}
\end{equation}
The relation between the induction components in the $xz$ plane is given by the system
\begin{align} 
	\label{EQn_106a_} 
	\left(\frac{\omega ^{2} }{c^{2} } -k^{2} \vartheta _{xx} \right)d_{x} 
		-k^{2} \vartheta _{xz}\; d_{z} =0 .
\end{align} 

Let us consider the propagation of waves without dissipation in the absence of external field. In this case Eq. \eqref{EQn_105_} leads to
\begin{align} 
&\left(\frac{\omega ^{2} }{k^{2} c^{2} } \varepsilon -1\right)
	\left[\omega ^{2} -\frac{E_{k}^{2} }{\hbar ^{2} } 
		+u\left(\theta \right)\frac{\varepsilon _{k} }{\hbar ^{2} } \right]= \notag\\ 
\label{EQn_107_} 
&\qquad=\eta \frac{\varepsilon +2}{6\varepsilon}\sin ^{2} \theta \cdot c^{2} k^{2} 
	+\eta \left(\frac{\omega ^{2} }{k^{2} c^{2} } \varepsilon -1\right)
		\left[\frac{2+\varepsilon}{6\varepsilon } 
			-\frac{\varepsilon -1}{2\varepsilon }\cos ^{2} \theta \right]\cdot c^{2} k^{2} , 
\end{align} 
where $u\left(\theta \right)=8\pi d_{0}^{2} n_{0} (1/3 -\cos^2 \theta)$ and $\varepsilon =1+4\pi n_{0} \alpha \left(\omega \right)$. Neglecting the interaction of the acoustic and electromagnetic waves, we can put $\eta =0$ in the right hand part of \eqref{EQn_107_}, and obtain the dispersion law for the condensate excitations, which is anisotropic and has the form
\begin{equation}
\omega ^{2} =\frac{k^{2} }{2m} \left\{\frac{\hbar ^{2} k^{2} }{2m} 
	+2n_{0} \Big[g+4\pi d_{0}^{2} \;\big(\cos ^{2} \theta - 1/3\big)\Big]\right\}. 
\label{EQn_108_}
\end{equation}

The phase velocity is
\begin{equation}
c_{d}^{2} 
	=c_{B}^{2} +\frac{4\pi n_{0} d_{0}^{2} }{m} \left(\cos ^{2} \theta -\frac{1}{3} \right), 
\label{EQn_109_}
\end{equation}
where $c_{B}^{2} =n_{0} g/m$ is the phase velocity in a system of Bose particles in a constant electric field $E_0$ without intrinsic dipole moment, when the dipole moment is induced by the field $d_{0} =\alpha _{0} E_{0} $ \cite{Brusov88}. In the first order by $\eta $, using $c_{d}^{2} \ll c^{2}$, the sound velocity can be brought to
\begin{equation}
c_{d}^{2} =c_{B}^{2} +\frac{4\pi n_{0} d_{0}^{2} }{m} \left(\cos ^{2} \theta -\frac{1}{3} \right)+\eta c^{2} \frac{\left(5-2\varepsilon \right)}{6\varepsilon } \cos ^{2} \theta . \label{EQn_110_}
\end{equation}
The last term in \eqref{EQn_110_}, which is the correction to the sound velocity due to coupling of the acoustic and electromagnetic waves, is of the same order as the second term. Sound propagation is accompanied by longitudinal oscillations of the electric field's intensity, such that 
\[\frac{b_{x}}{b_z}=\tan \theta .\]
The amplitudes of the field and condensate oscillations are related as
\begin{equation}
b_{z} =-\frac{2\pi \psi _{0} d_{0} }{\varepsilon } \Psi _{0} ,
	\qquad 
\frac{\Phi _{0} }{\Psi _{0} } =i\frac{\hbar \omega }{\varepsilon _{k} }. 
\label{EQn_111_}
\end{equation}

In the electromagnetic wave the intensity and displacement vectors lie in the $xz$ plane, and its phase velocity $c_{*} $ in the first order by $\eta $ is given by
\begin{equation}
\left(\frac{c_{*} }{c} \right)^{2} =\varepsilon ^{-1} 
	\left[1+\frac{\varepsilon +2}{6}\,\eta\sin ^{2} \theta \right],
\label{EQn_112_}
\end{equation}
while
\begin{align}
\frac{d_{z}}{d_x }&=-\tan\theta ; \label{EQn_113_}\\
\frac{b_{z} }{b_{x} } 
	&=-\left[1+\frac{\varepsilon+2}{6}\, \eta \right]\tan\theta . 
\label{EQn_114_}
\end{align}

Thus in the BEC of polar atoms the dipole-dipole interaction leads to the appearance of new term in the expression for the sound velocity \eqref{EQn_110_}, which depends on the angle between the wave vector and the dipole moment. The electromagnetic wave has two modes. One of them, with dispersion \eqref{EQn_103_}, has the electric intensity and displacement vectors oscillating normal to the plane of vectors $\mathbf{s}$ and $\mathbf{e}$, while the condensate is not excited. In the second mode the electric intensity and displacement vectors oscillate in the plane of vectors $\mathbf{s}$ and $\mathbf{e}$, so that the displacement vector is normal to the wave vector \eqref{EQn_113_}, and the intensity vector has the longitudinal component \eqref{EQn_114_}. The amplitudes of condensate oscillations, according to \eqref{EQn_90_}, \eqref{EQn_91_}, are given by
\begin{equation}
	\Psi _{0} =
		-\frac{\psi _{0} d_{0} }{mc^{2} } \varepsilon _{0} \left(\varepsilon _{0} +2\right)b_{z},
			\qquad 
	\Phi _{0} =\pm i\frac{2}{\sqrt{\varepsilon _{0} } } \frac{mc}{\hbar k} \Psi _{0} . 
\label{EQn_115_}
\end{equation}

\section{Conclusion}
In this paper we study the theory of propagation of electromagnetic and acoustic waves in a Bose-Einstein condensate of particles having an intrinsic dipole moment. We take into account the electric polarization of atoms by external electric field. The Bose-Einstein condensate is described by a modified Gross--Pitaevskii equation, which incorporates relaxation effects by introducing a phenomenological dissipative coefficient associated with the third coefficient of viscosity and the time of homogeneous relaxation of condensate particles' number density. Interaction of the condensate's atom with the electric field is considered in the dipole approximation. The considered medium is anisotropic due to existence of the preferred direction of spontaneous orientation of atomic dipole moments.

We derive the permittivity tensor of the BEC, in which the atoms interact via short-range forces, and show, that this medium has both temporal and spatial dispersion. We consider the propagation of both acoustic and electromagnetic waves in the case when the dispersion curves of the two modes do not intercross. It is shown that the propagation of sound waves in the condensate can be accompanied by oscillations of the electric field. Experimental observation of electric field oscillations in the wave of second sound in superfluid helium was reported in \cite{Ryb04}.

We also study the propagation of electromagnetic waves in the BEC possessing excitations with an energy gap. In this case the dispersion curves of acoustic and electromagnetic waves intercross, which leads to strong hybridization in the vicinity of the intersection point. This study was motivated by the observation of resonant absorption of microwave radiation in liquid helium \cite{Ryb09,Ryb10}, which was interpreted in \cite{YM14} as proof of the existence in superfluid helium, along with acoustic excitations, of quasiparticle excitations with the gap. The influence is considered of the dipole-dipole interaction between the condensate atoms on the propagation of both electromagnetic waves and acoustic excitations.

\end{document}